\documentclass[12pt]{article}
\usepackage{amsmath}

\usepackage{graphicx}
\usepackage{array}
\usepackage[top= 2.5cm,bottom= 2.5 cm, left=2cm,right=2cm]{geometry}

\begin{document}

\title{The Quark Dirac Sea and the Contracted Universe\\
Cooperate to Produce the Big Bang with the Quark Energy}
\author{Jiao-Lin Xu\\
$\qquad$$\qquad$$\qquad$$\qquad$$\qquad$$\qquad$\\
The Center for Simulational Physics,\\
The Department of Physics and Astronomy\\
University of Georgia,\\
Athens, GA 30602, USA\\
\texttt{jxu@hal.physast.uga.edu}\\
$\qquad$$\qquad$$\qquad$$\qquad$$\qquad$$\qquad$\\
Institute of Theoretical Physics,\\
 Chinese Academy of Sciences, P.
O. Box 2735, Beijing 100080, China}
\date{\today}
\maketitle
\begin{abstract} The Big Bang theory cannot and does
not provide any explanation for the primordial hot and dense initial
condition. In order to give an explanation for the cause of the Big
Bang, this paper expands the original Dirac sea (which includes only
electrons) to the quark Dirac sea (QDS) including quarks (u and d) for
producing the Big Bang with quark energy.

The QDS is composed of ``relatively infinite" u-quarks and d-quarks
as well as electrons with negative energy in the vacuum. A huge
number of domains with sizes much smaller than $10^{-18}$m of the
body-central cubic quark lattice with a lattice constant ``a" =
Planck length ($1.62\times10^{-35}m$) are distributed randomly over
the QDS. The QDS is a homogeneous, isotropic, equivalent ``continuous"
and ``empty" (no net electric charge, no net color charge, no
gravitational force field since the gravitational potential is the
same at any physical point in the QDS) perfect vacuum model.

The gravity of the universe pulls on the quarks inside the QDS. The
pulling force becomes larger and larger as the universe shrinks and
shrinks. Once the pulling force is larger than the binding force on
the quarks by the whole QDS, a huge number of quarks and antiquarks
will be excited out from the QDS. This is a necessary and sufficient
condition for the Big Bang.

The huge number of excited quark-antiquark pairs annihilate back
to the QDS and release a huge amount of energy; the huge number of
excited quarks (or antiquarks) combine into baryons (or antibaryons)
and release a huge amount of energy; the combined baryon-antibaryon
pairs annihilate back to the QDS and release a huge a mount of energy
also. Together, these huge quark energies make the Big Bang at its
finite size with radius R = 2.24$\times10^{4}$m.
\end{abstract}

\section{Introduction}
The Big Bang \cite{Big Bang} is the cosmological model of
the universe that is best supported by all
lines of scientific evidence and observation \cite{Hubble}
\cite{Microwave}. The essential idea is that the universe has
expanded from a primordial hot and dense initial condition at
some finite time in the past and continues to expand
to this day.

``Without any evidence associated with the earliest instant of
the expansion, the Big Bang theory cannot and does not provide
any explanation for such an initial condition; rather, it
describes and explains the general evolution of the universe since
that instant." (Big Bang-Wikipedia, the free encyclopedia).
This is a weak point of the Big Bang theory. Anyway we should give
a reasonable explanation for the initial condition (the origin) of
the Big Bang. This paper will spell out where the energy of the
Big Bang comes from.

All known energies, however, are not large enough to make such a
special anomalous Big Bang. Nuclear fusion energy is the highest
energy humans have ever known, it still cannot make the Big Bang.
For example, a large quantity of hydrogen has been fusing to helium
in our closest star, the Sun, for billions of years. Although these
nuclear fusion reactions produce large quantities of energy, it is
not enough to overcome the gravitation of the Sun. Thus the nuclear
fusion energy is not enough to make the Big Bang of the universe.
This paper suggests that there is one and only one kind of energy,
the quark energy, which can make the Big Bang.

In order to improve the Big Bang theory, this paper tries to provide
a mechanism to produce the Big Bang naturally using the quark energy.
Where does the quark energy comes from? There is only the vacuum which
might provide the quark energy and there is nothing else.

Inside the current vacuum model (the Dirac sea model), however, there
is only electrons that cannot provide the quark energy. Thus, we have
to expand the Dirac sea model from only including electrons to
including quarks (u and d) and electrons for providing quark energy from
the vacuum.

The Dirac sea \cite{Dirac} is a theoretical model of the vacuum as
an infinite sea of particles possessing negative energy. It was first
postulated by the British physicist Paul Dirac in 1930 to explain
the anomalous negative energy quantum states predicted by the Dirac
equation for relativistic electrons. At that time quarks were unknown,
so the Dirac sea only included electrons. Similarly, the quarks
\cite{Quark Model} are fermions with spin = $\frac{1}{2}$ and obey
the Dirac equation as well as have negative energy state solutions.
Therefore, from the original Dirac sea, we extend the model to a new
Dirac sea (the quark Dirac sea) which includes the quarks. Although
there are six flavored quarks, only two kinds of quarks (u and d) can
compose stable baryons. Since the vacuum is absolutely stable, as a
vacuum model, the new Dirac sea, cannot include other quarks (s, c,
b and t) which cannot compose any stable baryon. In fact the numbers
of these quarks (s, c, b and t) are too small and the lifetimes of
these quarks are too short to compose a stable vacuum. Thus we omit
them. In order to avoid confusion, we call the new Dirac sea,
including u-quarks and d-quarks as well as electrons, the quark
Dirac sea (QDS); we call the Dirac sea, including only electrons,
the original Dirac sea.

At the same time, we have to incorporate the advantages of the
current Big Bang models. There are many Big Bang models now.
They all have some advantages which we shall incorporate. Such as:

The inflationary model is to set up the initial conditions for the
standard model and explain the large-scale uniformity of the universe
\cite{Guth}. These arouse us. This paper will provides a mechanism
to produce a huge energy for the Big Bang and to produce a physical
firm foundation for the uniformity of the universe.

The oscillatory Big Bang model investigated by Einstein in 1930 and
critiqued by Richard Tolman \cite{Tolman} in 1934, in which the
universe undergoes a series of oscillations, each beginning with a
big bang and ending with a big crunch. The big crunch prepares for
the next Big Bang. We shall incorporate this.

The above theory has been revived in cosmology as the cyclic model
\cite{Baum} and the Big Bounce \cite{Bounce} \cite{Bojowald}, $\ldots$.

The ``primeval atom" model,  Lemaitre (1931) suggested that the
universe contrcted backward in time, and would continue to do so until
it could contract no further, bringing all the mass of the universe
into a single point, a ``primeval atom", before the point time and
space did not exist \cite{Lemaitre}. Hawking points suggested that
before the Big Bang, the universe contracted to a singularity with
infinite density and infinite temperature \cite{Hawking}. Although
we do not believe the ``primeval atom" really exist, we think that
before the Big Bang, the universe really contract to very small
size. We will incorporate this.
For
example, there is the oscillatory Big Bang model investigated by
Einstein in 1930 and critiqued by Richard Tolman \cite{Tolman} in
1934, in which the universe undergoes a series of oscillations, each
beginning with a big bang and ending with a big crunch. The theory
has been revived in cosmology as the cyclic model \cite{Baum} and the
Big Bounce \cite{Bounce} \cite{Bojowald}, $\ldots$. Using general
relativity, Hawking points out \cite{Hawking}: before the Big Bang,
the universe contracted to a singularity with infinite density and
infinite temperature (since quantum effects, the singularity can not
be reached). Although the above models are not completely the same,
they have a common characteristic in that the universe undergoes a
contraction process before the Big Bang. This is the common
characteristic of the Big Bang models, we shall incorporate this
into our mechanism.

We will briefly show how the contracted universe and the quark Dirac
sea can cooperate to produce the Big Bang with quark energy now.
According to the common characteristic of the above current Big Bang
models, the universe undergoes a contraction before the Big Bang.
Thus, the universe with a huge mass is compressed by the gravity of
the universe before the Big Bang. At the same time, the universe is
pulling on the quarks inside the quark Dirac sea (in the vacuum)
with gravitational force. As the size of the universe becomes
smaller and smaller, the pulling force becomes larger and larger.
Thus eventually a critical point will be reached: the universe's
gravitational pulling force (on the quarks) overcomes the quark
Dirac sea binding force (on the same quarks), and the quarks (and
their antiquarks) will be excited out from the quark Dirac sea
(the vacuum) negative energy state. Like electrons inside the Dirac
sea (the vacuum) negative energy state are no observable electric
charges and masses, but once they are excited out from the negative
state inside the Dirac sea, they obtain their observable electric
charges and masses. Similarly, although the quarks and antiquarks
are no observable electric charges, color charges and masses, once
the quarks and antiquarks are excited from the quark Dirac sea will
show their electric charges, color charges and masses. Thus the
exciting process increases the mass of the universe and the pulling
force (from the excited quarks and antiquarks) as well as decreases
the binding force of the quark Dirac sea. So that, more and more
quarks (and their antiquarks) are pulled out from the quark Dirac
sea. At the same time, the huge number of excited quarks (and their
antiquarks) will combine into baryons (antibaryons) and release a
huge amount of energy to produce the Big Bang. Three quarks combine
into a baryons to release much larger energy than a nuclear fusion
from the quark confinement fact. In addition, the baryon-antibaryon
pairs annihilate and release even more energy for the Big Bang. Once
the universe expands large enough, and the universe's pulling force
becomes small enough (smaller than binding force of the QDS), it
cannot pull the quarks (antiquarks) out any more. The universe has
already pulled out a huge amount of quarks and their antiquarks from
the quark Dirac sea. These quarks and antiquarks well produce a huge
amount of quark energy to produce the Big Bang. The details will be
shown in the following sections.

\section{The Original Dirac Sea--The Earliest Model of the Vacuum}The
earliest theoretical model of the vacuum is the Dirac sea as an
infinite sea of electrons possessing negative energy. It was invented
by the British physicist Paul Dirac in 1930 to explain the anomalous
negative-energy quantum states predicted by the Dirac equation for
relativistic electrons \cite{Dirac}. The positron, the antimatter
counterpart of the electron, was originally conceived as a hole in the
Dirac sea, well before its experimental discovery in 1932.
\subsection{The advantages of the original Dirac sea model}
The original Dirac sea theory makes three great predictions:

1. It predicts an antiparticle---a hole of the negative energy particle
in the Dirac sea.

2. In the original Dirac sea theory, the particle and its
antiparticle are not absolutely symmetric. The particles have
positive energy and outside the original Dirac sea; the antiparticles
are the holes of the original Dirac sea. Although the holes have an
equivalent electric charge and mass of the antiparticles, they are
temporary particles. Once they depart from the original Dirac sea,
they will disappear. Thus we can understand why there are only the
particle occupying the current universe.

3. It predicts that the empty space (the vacuum) is not really empty.
It is fully filled by the electrons with negative energy. Thus it can
produce electron-antielectron pairs (an electron is excited from the
original Dirac sea and leaves an electron hole in the original Dirac
sea) and it can accept electron-antielectron pair annihilations (the
electrons fill back into the holes). The original Dirac sea provides
a physical foundation for electron-antielectron pair production and
annihilation. If the original Dirac sea is not in the vacuum, it will
be difficult to understand why electron-antielectron pairs can be
produced from the vacuum (``the empty space").                                 .
\subsection{The weaknesses of the original Dirac sea theory}
Despite its success, at the same time, there are some weaknesses in
the original Dirac sea theory:

1. The existence of the original Dirac sea implies infinite
negative electric charges filling all of space. In order to make any
sense out of this, one must assume that the ``bare vacuum" must have
an infinite positive charge which is exactly canceled by the original
Dirac sea. We need an electrically neutral Dirac sea. Thus we have to
add particles with positive electric charges (such as the u-quarks
with Q = $\frac{2}{3}$) into the sea to cancel the negative electric
charges of the electrons.

2. The existence of the original Dirac sea implies infinite negative
electric charges filling all of space. Because of the repulsive
forces of the infinite negative electric charges, the original Dirac
sea (vacuum) will be unstable. We need a stable structure of the Dirac
sea. According to the solid state physics, the most stable structures
of many body systems are the lattices. Maybe an ideal Dirac sea will
have a lattice structure also.

3. Pauli exclusion does not definitively mean that a filled Dirac sea
can not accept more electrons. Since, as Hilbert elucidated, a sea of
infinite extent can accept new particles even if it is filled. We might
need another condition to ensure that the Dirac sea can not accept more
``electrons". What is the condition? Maybe a new model is needed.

4. Although the original Dirac sea theory has successfully explained
the electron-antielectron pair productions (electrons are excited from
the original Dirac sea and leave their holes in the sea) and
annihilations (the electrons go back into their holes in the original
dirac sea), it cannot explain the p-$\bar{p}$ pair and n-$\bar{n}$ pair
productions and annihilations since there is not any quark in the
original Dirac sea. These pair productions (from the vacuum) and
annihilates (back to the vacuum) are supported by a lot of experimental
facts. As a model of vacuum, it has to give a reasonable explanation.
Thus it is necessary to have u-quarks and d-quarks inside the Dirac sea.

The above weaknesses must be corrected. Thus we need a new model
of the vacuum to correct the weaknesses, and to incorporate and develop
the advantages of the original Dirac sea theory.

\section{The Quark Dirac Sea Model (QDS) of the Vacuum }
As mentioned above, the earliest model of the vacuum, the original Dirac
sea model, has three advantages. We must adopt and develop these
advantages. At the same time, there are some weaknesses in the original
Dirac sea model. We have to correct the weaknesses using a new model.
Since the birth of the original Dirac sea model, more than seventy-five
years have passed; during this period physics and astronomy have made
great progress. The new achievements of physics, especial the quark
model and the solid state physics, provide good conditions for
further improvement and development of the original Dirac sea model.
\subsection{The fundamental postulates of the quark Dirac sea model}
The quark model is one of the most important new achievements \cite
{Quark Model}. There are six flavored quarks in the model: u-quarks,
d-quarks, s-quarks, c-quarks, b-quarks and t-quarks. They are fermions
with spin s = $\frac{1}{2}$. They all obey the Dirac equation.
Like electrons, there are positive energy state solutions and negative
energy state solutions of the Dirac equation for the u-quarks, the
d-quarks, $\ldots$, in real wold. For the electrons there is an infinite
original Dirac sea of electrons with negative energy in the vacuum.
Similarly, there are an u-quark Dirac sea of the u-quarks with negative
energy, a d-quark Dirac sea of the d-quarks with negative energy,
$\ldots$, in the vacuum also. The numbers of the s-quarks, the c-quarks,
the b-quarksm and the t-quarks are much smaller than the numbers of the
u-quarks and the d-quarks. And only u-quarks and d-quarks can compose
stable matter. We do not believe that the unstable quarks (s, c, b and t)
can compose the absolutely stable vacuum. As an approximation, we assume
that the u-quarks and the d-quarks and the electrons mainly compose the
physical vacuum.

Although there are photons between the electrons, they are completely
neglected in the original Dirac sea model and various solid state lattice
models. Similarly, although there are gluons between the quarks, in order
to simplify, they will be phenomenologically neglected in the quark
Dirac sea model.

In order to avoid confusion with the mathematical and philosophic
term ``infinite", we define a ``relatively infinite" system as follows:
if a system is beyond the region of electric force and gravitational
force as well as human observation, it is ``relatively infinitely" far;
if the boundary of a system is as far as ``relatively infinite", we call
the system as ``relatively infinitely" large. In fact, some
``relatively infinitely" large systems are finite in mathematics and
in philosophy.

In order to make a phenomenological vacuum model, we assume:

\textbf{Postulate I: There is a ``relatively infinite" large quark
Dirac sea (QDS) which is completely filled with the negative energy
u-quarks and d-quarks as well as electrons inside the vacuum.}

According to Dirac sea theory \cite{Dirac}, the particles with negative
energy are in the vacuum state and there is not any observable physical
effect for these negative energy electrons. Similarly, the negative
energy u-quarks and d-quarks are in the vacuum state of the quark Dirac
sea, there is not any observable physical effect for these negative
energy quarks (u and d) also.

A huge system usually will be divided many parts. For example, a large
chunk of ferrous metal is always composed of many very small
micro-crystal grains. The ``relatively infinite" quark Dirac sea will
be divided into a huge number of domains also. The domains have not be
discovered by the standard model. Thus we guess that the sizes of the
domains are much smaller than the distance scale $10^{-18}$ m of the
standard model \cite{Standard}.

\textbf{Postulate II: The quark Dirac sea is fully filled with domains.
The domains are completely randomly distributed in sizes and in
positions as well as in directions of symmetry axes of the quark
lattice in the domains (see \textbf{Postulate III}). The sizes of the
domains are much smaller than the distance scalar $10^{-18}$ m of the
standard model.}

The domains are completely randomly distributed in sizes and in
positions as well as in directions of symmetry axis of the domains
(\textbf{see Postulate III}). Their sizes are much smaller than the
distance scale $10^{-18}$ m of the standard model. Thus, the quark
Dirac sea is homogeneous and isotropic as well as ``continuous" from
the statistics and the opinion of the standard model.

If there is an empty position of u-quark (or d-quark or electron) in
any domain, an u-quark (or d-quark or electron) with positive energy
will fall down into the position to make the domain completely full.

There are negative energy u-quarks and d-quarks as well as electrons
inside a domain. There are strong interactions and electromagnetic
interactions among these negative particles. The strong interactions
is much larger than the electromagnetic interactions, it determines
and holds the structure of the domains. The electrons are much smaller
than the quarks and they do not have strong interactions with the
quarks and the electrons. The electrons do not play an important role
in the structure of the domains. We only consider u-quarks and d-quarks,
while neglecting the electrons. Thus we think temporally there are only
the u-quarks and the d-quarks in the domains. The strong pulling forces
drive the u-quarks and the d-quarks to form the densest quark Dirac
sea in the vacuums.

From the experimental mass (938 Mev) of proton (uud) is close to the
mass (940 Mev) of neutron (udd), we can estimate that the mass of u
is close to the mass of d and the radius $r_u$ of the u-quarks is
close to the radius $r_d$ of the d-quarks:
\begin{equation}
r_u\,\,\approx\,\, r_d. \label{Ru and Rd}
\end{equation}

If using the current quark masses: $m_u$ = 2.55 Mev and $m_d$ = 5.04
Mev \cite{Quark mass}, and assume their densities are the same
($\rho_u$ = $\rho_d$), the volume $V_u$ = $\frac{m_u}{\rho_u}$ and
the volume $V_d$ = $\frac{m_d}{\rho_d}$. Using the Formula V =
$\frac{4\pi r^3}{3}$ $\to$ r = $\sqrt[3]{\frac{3V}{4\pi}}$, we have
\begin{equation}
\frac{r_u}{r_d}\,=\,\sqrt[3]{\frac{m_u}{\rho_u}}\div
\sqrt[3]{\frac{m_d}{\rho_d}}\,= 0.797
\label{ru/rd}
\end{equation}
From both Formulae (\ref{Ru and Rd}) and (\ref{ru/rd}), we get
\begin{equation}
1\,\, \geq \,\, r_u/r_d \,\,\geq \,\, 0.73, \label{ru/rd}
\end{equation}

According to the crystal structure theory \cite{Crystal}, for two
kinds of different particles (such as u and d), if 1 $\geq$ $r_u$/$r_d$
$\geq$ 0.73, the densest structure of the u-quarks and the d-quarks is a
body-central cubic quark lattice. Note that inside the quark lattice,
the eight nearest neighbors of any u-quark are all d-quarks and the
eight nearest neighbors of any d-quark are all u-quarks.

The u-quarks have electric charge $Q_u$ = +$\frac{2}{3}$ and the
d-quarks have electric charge $Q_d$ = -$\frac{1}{3}$. Coulomb's
repulsive forces separate the u-quarks with the same charges and the
d-quarks with the same charges; while Coulomb's attractive forces pull
the u-quarks and the d-quarks to approach each other since they have
positive and negative charges. So that the u-quarks and the d-quarks
form a quark lattice that the nearest neighbors of any u-quark are all
d-quarks and the nearest neighbors of any d-quark are all u-quarks. The
body-central cubic lattice is just this kind of lattice.

Thus, the strong interactions and electromagnetic interactions of the
u-quarks and the d-quarks cooperate to form a super strong body-central
cubic quark lattice as shown in Figure 1 (a).

If we consider the u-quarks in the quark Dirac sea only, the u-quarks
form a simple cubic u-quark lattice (see Figure 1 (b)). If only consider
the d-quarks inside the quark lattice, they form a simple cubic d-quark
lattice also (see Figure 1 (c)).

Figure 1. The body-central cubic quark lattice. Figure 1 (a) shows the
body-central cubic quark (u and d) lattice. The u-quarks are at the
corners of the conventional cells, the d-quarks are at the centers of
the conventional cells. Figure 1 (b) shows the simple cubic u-quark
lattice. Figure 1 (c) shows the simple cubic d-quark lattice.
\,\,\,\,\,\,\,\,\,\,\,\\
\includegraphics[scale=0.80,angle= 0]{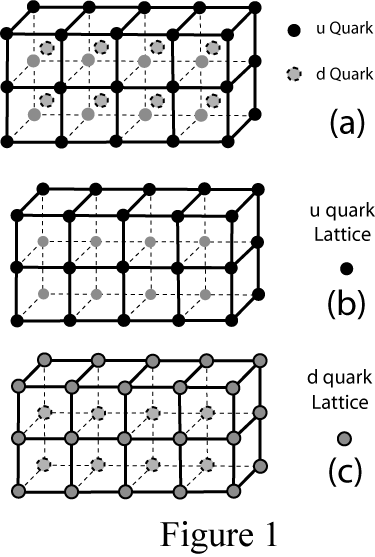}\\

In the lattice, a kind of quarks (such as d-quarks) are at the centers
of the conventional cells and the other kind of quarks (u-quarks) are
at the corners of the conventional cells. The conventional cell and the
primitive cell are shown in Figure 2. In Figure 2 (b), the d-quark (at
the center) and the u-quarks (at the corners). The center quarks and
the corner quarks can exchange with each other. This depends
on your chosen coordinates (the origin point on which quark).\\
\,\,\,\,\,\,\,\,\\ \includegraphics[scale=0.5]{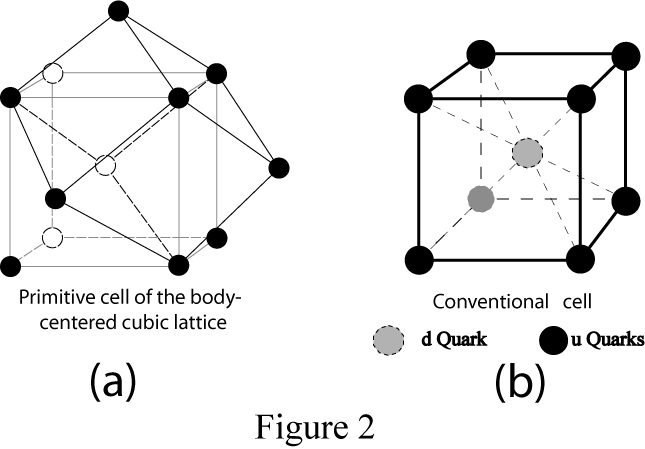}

Figure 2. The primitive cell and the conventional cell of the
body-central cubic quark lattice. Figure 2 (a) shows a primitive cell.
The primitive cell shown is a rhombohedron of edge $\frac{\sqrt{3}a}{2}$
and the angle between adjacent edges is $109^{\circ} 28^{'}$.
Figure 2 (b) shows a conventional cell. A d-quark at the center and
eight u-quarks are at the eight corners of the conventional cell.

There is a piece of body central cubic quark (u and d) lattice
in a domain of the quark Dirac sea. The sizes of the domains are
much smaller than the distance scale $10^{-18}$ m of the standard
model since the standard model has not found the domains and the
standard model works at distance scalar $\geq$ $10^{-18}$ m
\cite{Standard}. The periodic constant ``a" is very important.
First, it has to be much smaller than the size of the domains (the
sizes of domains are much smaller than $10^{-18}$m), i.e. ``a"
$\ll$ $10^{-18}$m. Second, loop quantum gravity point out
that the space is a mesh of tiny ``atoms" (spheres). The diameter of
the atoms is the Planck length \cite{Bojowald}. Thus we take the
lattice constant as the Planck length `a' = 1.62$\times10^{-35}$m
\cite{Planck length}.

\textbf{Postulate III: \textbf{Each domain is a piece of
body-central cubic quark lattice. In the lattice, one kind of quarks
(such as d) are at the centers, the other kind of quarks (such as u)
are at the corners, and the electrons are freely moving inside the
lattice. The quark lattice constant is the Planck length ``a" =
1.62$\times10^{-35}$ m.}}

The physical vacuum has no net electric charge. Thus the number of
u-quarks ($N_u$) equals the number of the d-quarks ($N_d$) equals
3 times the number of the electrons ($N_e$) in the perfect
quark Dirac sea domains (3$N_u$ = 3$N_d$ = $N_e$) to ensure the
total electric charge of the quark Dirac sea equals zero in each
domain. Each primitive cell of the body-central cubic lattice has
two quarks \cite{Primitive cell}, the u-quark with $Q_u =\,+
\frac{2}{3}$ and the d-quark with $Q_d = \,-\,\frac{1}{3}$, it has
total Q = +$\frac{1}{3}$. Each of the three primitive cells add an
electron with $Q_e$ = -1. Thus the three primitive cells have electric
charge Q = 0. For any physical point, we mean that a small ball with
radius r = $10^{-18}$ m and its center at the point, and the total
electric charge $Q_{Point}$ = 0.

\textbf{Postulate IV: The number of the u-quarks equals the number of
the d-quarks equals three times the number of the electrons in each
domain of the quark Dirac sea. Each primitive cell has two quarks (u
and d), nearest three primitive cells contain an electron.}

The physical vacuum has no net color charge. Thus, the number of
u-quarks with red color equals the number of u-quarks with green
color equals the number of u-quarks with blue color, and the
number of d-quarks with red color equals the number of d-quarks with
green color equals the number of d-quarks with blue color, to ensure
the total net color charges of the quark Dirac sea domain equals zero.
Each primitive cell has different colors for each quark, three nearest
neighboring primitive cells have white color for the quark-u and the
quark-d.

\textbf{Postulate V: For the u-quarks and the d-quarks there are the
same numbers of red quarks and green quarks as well as blue quarks in
each domain of the quark Dirac sea. Three nearest neighboring
primitive cells have white color (or colorless).}

The quark Dirac sea has ``relatively infinite" number of u-quarks and
d-quarks as well as electrons in the vacuum. Since these particles
all have finite masses, the quark Dirac sea has ``relatively infinite"
mass. Because the quark Dirac sea is homogenous and isotropic, the
gravitational potential is homogenous and isotropic (a constant) in
the vacuum. The gravitational potential is a constant anywhere yields
that the gravitational field equals zero at anywhere in the quark
Dirac sea.

\textbf{Postulate VI: The quark Dirac sea provides a constant
gravitational potential $V_{QD}$ (maybe ``relatively infinite") at
any point. We select the potential as the energy zero point $V_{QD}$
= 0. The constant gravitational potential leads to the gravitational
field $F_G$ equals zero ($F_G$ = 0).}

The quark Dirac sea is a homogeneous, isotropic, ``continuous" and
``empty" (no electric charge, no color charge, no effective ``mass"
(no gravitational potential and no gravitational field) at any
physical point in the quark Dirac sea) perfect vacuum model.

\subsection{The quark Dirac sea model (QDS) corrects the three
weaknesses of the original Dirac sea model}
1. The original Dirac sea model implies an infinite negative
electric charges filling all of the space. This is not in
agreement with the fact that the vacuum is without any electric
charge. The new model (QDS) has no electric charge and color
charge at any physical point of the quark Dirac sea. Thus it is
in agreement with the empty vacuum space.

2. The original Dirac sea is unstable since there are the forces of
repulsion between the negative electric charges of the electrons in
the sea. For the quark Dirac sea model, however, there are strong
pulling forces between the quarks (no strong repulsion). In the QDS,
for any u-quark with Q = +$\frac{2}{3}$, there are always eight
nearest neighbor d-quarks with Q = -$\frac{1}{3}$; for any d-quark
with Q = -$\frac{1}{3}$, there are always eight neighbors u-quarks.
Thus the pulling electric forces between the quarks dominate over the
repulsions. The quark Dirac sea is a super-stable structure, it is
real like the vacuum space that has never been changed.

3. The original Dirac sea might be, as Hilbert elucidated
(Hilbert's paradox of the Grand Hotel), a sea of infinite
extent that can accept new particles even if it is filled.
The quark Dirac sea is composed of domains. Any domain is
fully occupied by the finite u-quarks and the finite d-quarks
as well as finite electrons, and there is not any empty position
from \textbf{Postulate II}. According to Pauli exclusion, there
is zero possibility to accept any particle in any domain. The
possibility of any domain accepting any particle is zero. Since
infinite zero sum together is still zero ($\sum0$ = 0), the ideal
quark Dirac sea cannot accept any additional particles.
\subsection{The quark Dirac sea (QDS) incorporates and develops
the advantages of the original Dirac Sea}
The quark Dirac sea model not only corrects all three weaknesses
of the original Dirac sea, but also incorporates and develops
all advantages of the original Dirac sea:

1. The quark Dirac sea predicts the antiparticles, these are the holes
in the quark Dirac sea: An electron hole of the quark Dirac sea is an
antielectron, an u-quark hole of the quark Dirac sea is an anti-u-
quark and a d-quark hole of the quark Dirac sea is an anti-d-quark.
The quark Dirac sea not only provides a natural explanation of the
antielectron, but also produces a natural explanation of anti-u-quarks
and anti-d-quarks.

2. The quark Dirac sea predicts that the particle-antiparticle pairs
usually are created simulteteneously or annihilate with each other.
A quark (u or d) is excited from the quark Dirac sea and an antiquark
($\bar{u}$ or $\bar{d}$) left in the quark Dirac sea at the same time.
This is a quark-antiquark pair production. On the other hand, when a
quark (u or d) falls into its hole (u-hole or d-hole) in the quark
Dirac sea, this is a quark-antiquark pair annihilation. Similarly, if
three quarks (uud or udd inside the QDS) are hit by the same force at
the same time with the similar strongth, they may be excited together
from the quark Dirac sea, a three fold hole ($\overline{uud}$ or
$\overline{udd}$) will be left in the quark Dirac sea, this is a
baryon-antibaryon production. A baryon (uud or udd) falls back its three
fold hole ($\overline{uud}$ or $\overline{udd}$) this is a baryon-
antibaryon pair annihilation. The quark Dirac sea not only can
explain the quark pairs (u$\bar{u}$ or d$\bar{d}$) production and
annihilation, but also can explain the baryon-antibaryon
(uud$\overline{uud}$, or udd$\overline{udd}$) pair production and
annihilation.

3. The quark Dirac sea predicts that the ``empty" space (the vacuum)
is not really empty. The ``empty" space is fully occupied by u-quarks
and d-quarks as well as electrons with negative energy. These quarks
and electrons form a super-strong homogeneous and isotropic structure,
and have no electric charge, no color charge and no effective ``mass"
(gravitational potential $V_{QD}$ = 0 and field $F_G$ = 0$F_G$ = 0
from \textbf{Postulate VI}) at any physical point ``empty"
vacuum. This provides a perfect super-stable stage for various
physical bodies, particles, stars and fields.
\subsection{The quark Dirac sea (QDS) is a perfect vacuum model}
According to the original Dirac sea theory \cite{Dirac}, the infinite
electrons with negative energy make the vacuum background without
any observable effect. Similarly, the negative energy u-quarks and
d-quarks as well as electrons in the quark Dirac sea form a perfect
physical vacuum without any observable effect also. This is true
due to four very strong reasons:

First, from \textbf{Postulate II}, the domains are completely randomly
distributed in positions and in directions of symmetry axis of the
body-central quark lattice in the domains. Thus, the quark Dirac Sea
is homogeneous and isotropic from statistics. The domain sizes are much
smaller than the distance scale $10^{-18}$ m of the standard model. The
lattice constant ``a" = 1.62$\times10^{-35}$ m is much smaller than the
distance scale $10^{-18}$ m of the standard model. Therefore the standard
model looks at the quark Dirac sea as a homogeneous and isotropic
as well as ``continuous" space.

Second, there are no electric charge and no color charge as well as no
effective ``mass" (no gravitational potential and field from \textbf{
Postulate VI}) at any physical point (a ball with radius = $10^{-18}$ m)
inside the QDS. Thus the quark Dirac sea appears as an empty space.

Third, the quark Dirac Sea is a super-strong structure (from Formula
(\ref{U(Tol)}) a quark is bound by - 4.03$\times10^9$ J energy).
All known phenomena cannot change the quark Dirac Sea since their
energies are too small. For example, the most highest energy phenomena,
nuclei fusions, only have energy about 10 Mev = $10^{-12}$J $\ll$ 4.03
$\times10^9$ J. Thus the quark Dirac sea seems to be an unchangeable
station of all particles and physical bodies as well as stars $\ldots$.
It plays the same role as the vacuum.

Fourth, when a particle is moving in ``empty" space, in fact, it is
moving in the quark Dirac sea since the space is fully filled with the
quark Dirac sea. The particle, in fact, is really moving in the periodic
field of the QDS. According to quantum mechanics \cite{QM}, an ideal
lattice does not scatter the particle moving inside the lattice. If the
Hamiltonian H of the particle moving in the lattice is independent from
time, the states of the particle are stationary. For such a state, the
expectative value of any physical quantity which does not depend on
time explicitly is constant. Thus a particle is moving in an ideal
periodic field of the quark Dirac sea, the ideal periodic field makes
the Hamiltonian to be independent from time. Hence the states of the
particle are stationary. So that, the particle looks like to be
moving in the vacuum.

The four reasons show that the quark Dirac sea is a homogeneous,
isotropic, ``continuous" and ``empty" space. The particles are moving
in the quark Dirac sea as if they are moving in a stable and
non-scattering space. Therefore the negative energy particle in the
quark Dirac sea do not have any observable effect in the usual
phenomena. The quark Dirac sea, however, will play a critical role
in the Big Bang of the universe as shown in the following sections.

\section{A Huge Number of Quarks and Antiquarks are Pulled Out from
the Quark Dirac Sea by the Gravity of the Contracted Universe}
It is a necessary and sufficient condition for the Big Bang that the
contracted universe pulls a huge number of quarks and antiquarks from
the quark Dirac sea. First, we find the pulling energy of a quark
inside the quark Dirac sea by the contracted universe.
\subsection{The gravitational pulling energy of a quark in the QDS
by the contracted universe}
According to the above (Introduction), a common characteristic of the
current Big Bang models (the oscillation, the cyclic model and the Big
Bounce, $\ldots$, etc) is that the universe undergoes a series of cycles,
each beginning with a Big Bang and ending with a big contraction, before
(or after) a Big Bang, the universe will contract to a small size. We
assume that the universe has shrunk to a small ball with radius R and mass
$M_U$ as well as a density $\rho(r,\theta,\phi)$ = $\frac{c}{r}$. The
universe's contraction is very slow because of the resistances from high
density and temperature of matter in the contracted universe. Thus we can
use Newtonian mechanics to estimate the pulling energy to a quark inside
the QDS by the contracted universe. The pulling energy is defined as the
potential energy between the quark (inside the QDS) and the whole
contracted universe. A spherical polar coordinate system (r, $\theta$,
$\phi$) is shown in Figure 3.\\
\,\,\,\,\,\,\,\,\,\,\,\,\,\,\,\,\,\,\,\,\,\,\,\,\\
\includegraphics[scale=0.53]{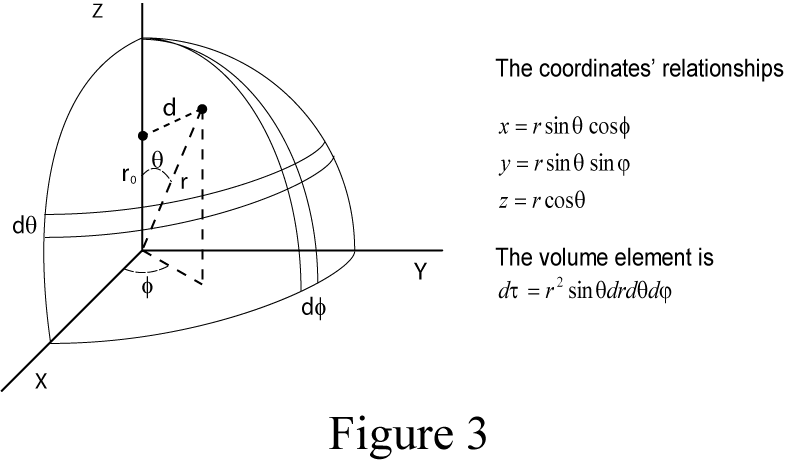}

Figure 3. A spherical polar coordinate and a rectangular cartesian
coordinates. The relationships of the two coordinates have been shown on
Figure 3. The coordinates of the quark are (r = $r_0$, $\theta$ = 0,
$\phi$ = 0) and for any point the coordinates are (r, $\theta$,
$\phi$).

Using spherical polar coordinates, for any point at (r, $\theta$, and
$\phi$), assuming the mass density of the universe is $\rho$(r, $\theta$,
$\phi$) = $\frac{c}{r}$ (c is a constant), the mass $M_U$ of the universe
is
\begin{equation}
M_U\,\,= \,\,\int\int\int \frac{c}{r}\,r^{2}sin\theta dr d\theta
d\phi\,= 2\pi c R^2, \to\,\, c = \frac{M_U}{2\pi R^2}.
\label{density}
\end{equation}

We calculate the gravity pulling energy $U(r_0,\theta,\phi)$ of a
quark inside the quark Dirac sea by the whole contracted universe
with mass $M_U$ and radius R now. The potential (pulling) energy
u(r, $\theta$, $\phi$) of the quark at the point with coordinates
(r = $r_0$, $\theta$=0,$\phi$=0) inside the quark Dirac sea
(negative energy state in the vacuum) pulled by a small mass piece
$\rho$(r, $\theta$, $\phi$) of the universe at (r, $\theta$, $\phi$)
is
\begin{equation}
u(r,\theta,\phi)\,\,\, = \,-\,\frac{Gm_q \rho(r)}{d} \,\,\,=
\,-\, \frac{Gm_q\,c\,} {r\sqrt{r^2+r_0^2-2r_0rcos\theta}},
\label{u(q)}
\end{equation}
where G = 6.6742$\times$$10^{-11}m^3kg^{-1}s^{-2}$, d is the
distance between any point (r, $\theta$, $\phi$) and the point
that the quark occupies ($r = r_0$, $\theta$ = 0, $\phi$ = 0);
the quark mass $m_q$ is the average mass of the mass $m_u$ of
the u-quark and the mass $m_d$ of the d-quark. The $m_u$ = 2.55 and
the $m_d$ = 5.04 \cite{Quark mass}, thus $m_q$ = 3.80 Mev.

Using Formula (\ref{u(q)}), we can calculate the pulling (the
gravitational potential) energy $U(r_0, \theta, \phi)$ of a quark
that is pulled by the whole universe as follows:
\begin{equation}
U(r_0,\theta,\phi) = -\int\int\int u(r,\theta,\phi)r^2\,sin\theta
dr d\theta d\phi = -\int\int\int\frac{Gm_qc}{r\sqrt{r^2+r_0^2-
2r_0rcos\theta}} r^2\,sin\theta dr d\theta d\phi. \label{mM0}
\end{equation}
Using Sin$\theta$d$\theta$ = d(-Cos$\theta$) and assuming -Cos$\theta$
= t, when $\theta$ = 0, t = -1 and $\theta$ = $\pi$, t = 1.
Letting $r^2+r_0^2$= b(r) and $2r_0r$ = a(r), from (\ref{mM0}), we
can get
\begin{equation}
U(r_0, \theta, \phi) = -\int\int\int\frac{Gm_qc}{\sqrt{a(r)t+b(r)}}
\,r dt dr d\phi. \label{mM1}
\end{equation}
Then using integrating formula $\int\frac{1}{\sqrt{at+b}}$dt =
$\frac{2}{a}\sqrt{at+b}$, we have
\begin{equation}
U(r_0,\theta,\phi) = - \int\int (Gm_qc)[\frac{2}{a(r)}\sqrt{a(r)t+b(r)}]
^{t=1}_{t=-1}\,r dr d\phi. \label{mM2}
\end{equation}
Putting a(r) = $2r_0r$ and b(r) = $r^2+r_0^2$ in the above formula and
making sure when t = -1, $\sqrt{a(r)t+b(r)}$ to get real values.
When t = -1, $\sqrt{a(r)t+b(r)}$ = $\sqrt{+r_0^2-2r_0r+r^2}$.
For r $\geq$ $r_0$, $\sqrt{a(r)t+b(r)}$ = (r - $r_0$); for r $\leq$
$r_0$, $\sqrt{a(r)t+b(r)}$ = $r_0$ - r. From (\ref{mM2}), we have
\begin{equation}
U(r_0,\theta,\phi)= - 2\pi Gm_q c[\int_{r_0}^R\frac{2r}{2r_0r}[(r+r_0)
-(r-r_0)]dr+\int_{0}^{r_0}\frac{2r}{2r_0r}[(r+r_0)-(r_0-r)]dr].
\label{mM3}
\end{equation}

From (\ref{density}) c = $\frac{M_U}{2\pi R^2}$, putting it in the above
expressed formula, we get
\begin{equation}
U(r_0,\theta,\phi) = -4\pi Gm_q c \frac{1}{r_0}[r_0(R-r_0)+
\frac{(r_0)^2}{2}]= - 4\pi Gm_q c (R-\frac{r_0}{2}) =
- \frac{2GM_U m_q}{R}(1-\frac{r_0}{2R})
\label{U(r0)}
\end{equation}

Formula (\ref{U(r0)}) shows that the universe's pulling energy is
independent from $\theta$ and $\phi$ as well as r. It is only dependent
on $r_0$. At the center of the universe ball, $r_0$ = 0, there is the
maximum pulling potential energy (absolute value). From Formula
(\ref{U(r0)}), the energy U(cent) is
\begin{equation}
U(cent)\,\,\,= \,-\,\frac{2GM_{U}m_{q}}{R}. \label{U(cent)}
\end{equation}
Also from Formula (\ref{U(r0)}), the pulling energy U(surf) of a quark
at the universe surface's ($r_0$ = R) is
\begin{equation}
U_{surf} =  -\,\frac{GM_{U}m_{q}}{R} \label{Usurf}
\end{equation}
Again from Formula (\ref{U(r0)}), using $m_{b}$ of the baryon mass instead
the quark mass $m_{q}$, we can get the pulling energy U(b,U,$r_0$) of a
baryon at the $r_0$ by the universe and the pulling energy U(b,U) of a
baryon at the center of the universe by the universe:
\begin{equation}
U(b,U,r_0)\,\,= \,-\,\frac{2GM_{U}m_{b}}{R}(1-\frac{r_0}{2R})\,\,\,\,\,
and\,\,\,\,\,U(b,U)\,\,= \,-\,\frac{2GM_{U}m_{b}}{R}. \label{U(b,U)}
\end{equation}

\subsection{The contraction of the universe cannot make
the Big Bang}
The contraction of the universe indeed can make a high density and a
high temperature. For example, when the universe contracts to a small
ball with R = $2.24\times10^4$ m (the critical R value from (\ref{R})),
if we know the observable mass $M_U$ of the current universe we can
find the average density $\bar{\rho}$ = $\frac{3M_U}{4\pi R^3}$ of
the universe. The measured mass value of the universe is different
using different methods. Such as:

A. K. Velan (1992), $M_U$ = 5.86$\times10^{53}$kg \cite{MU1};

J. Bernstein (1995), $M_U$ = the number of protons in the visible
universe $\times$ the mass of proton = $10^{77}\times1.67\times
10^{-27}$kg = 1.67$\times10^{50}$kg \cite{MU2};

H. Kragh (1999), $M_U$ = $\frac{c^3}{2GH}$ = $\frac{c^2}{2G}\times
\frac{c}{H}$ = $\frac{c^2}{2G}\times1.2\times10^{26}$ \cite{Planck
length} = 0.81$\times10^{53}$kg \cite{MU3};

D. Harrison (2000), $M_U$ = a mass of $10^{80}$ nucleons = $10^{80}\times1.67\times10^{-27}$kg =
1.67$\times10^{53}$kg \cite{MU4}.

From these mass values, we roughly estimate that
\begin{equation}
M_{U} \approx 1.0\times10^{53} kg. \label{Universe}
\end{equation}

Then the average density $\overline{\rho}$ of the universe with
R = $2.24\times 10^4$ m is
\begin{equation}
\overline{\rho}\,\,= \frac{1.00\times10^{53}kg}{\frac{4\pi}{3}
(2.24\times10^4m)^3}\,= \,2.13\times10^{39}(kg/m^3). \label{rho}
\end{equation}

There is not a formula with which one can deduce the temperature of
the contracted universe center. In order to estimate the temperature,
we need a phenomenological temperature formula. We assume that the
temperature of a celestial sphere center is proportional to its mass
and $\frac{1}{R}$ (R is the radius of the sphere):
\begin{equation}
T_s = C_s M_s \frac{1}{R_s} \label{TMRs}
\end{equation}
Using Formula (\ref{TMRs}) to the earth, we have
\begin{equation}
T_{Earth} = C_{Earth} M_{Earth} \frac{1}{R_{Earth}}. \label{Earth}
\end{equation}
Using Formula (\ref{TMRs}) to the Sun, we have
\begin{equation}
T_{Sun} = C_{Sun} M_{Sun} \frac{1}{R_{Sun}}. \label{Sun}
\end{equation}
From Formulae (\ref{Sun}) and (\ref{Earth}), we can get
\begin{equation}
T_{Sun}/T_{Earth} = C \frac{M_{Sun}}{M_{Earth}} \frac{R_{Earth}}
{R_{Sun}}, \label{Sun-Earth}
\end{equation}
where C = $\frac{C_{Sun}}{C_{Earth}}$. From (\ref{Sun-Earth}), we get
the temperature of the Sun center is
\begin{equation}
T_{Sun}\,=\,C\frac{M_{Sun}}{M_{earth}}\frac{R_{earth}}{R_{Sun}}
T_{earth}, \label{Tsun}
\end{equation}
Using $M_{Sun}$ = 1.99$\times10^{30}$kg, $M_{earth}$ = 5.97
$\times10^{24}$kg, $R_{Sun}$ = 6.961$\times10^8$m,
$R_{earth}$ = 6.38$\times10^3$m and $T_{earth}$ =  6000 K \cite{Earth T}
(there are many different published values for the temperature, we think
this value is reasonable). There are many different temperature values of
the Sun. We use the values $T_{Sun}$ = $15.7\times10^{6}$K of the solar
center \cite{Sun T}. From Formula (\ref{Tsun}), we get the constant C =
8.58$\times10^2$. We can use Formula (\ref{Tsun}) as a phenomenological
temperature formula of the universe center to estimate the temperature
of the contracted universe center.

Using $M_U$ = 1.0$\times 10^{53}$kg from (\ref{Universe}), then
$M_{earth}$ is the mass of the earth, and $R_{earth}$ is the radius of
the earth. In order to get useful values, we assume the radius of the
contracted universe $R_U$ = 2.24 $\times10^{4}$m, the temperature $T_U$
of the center of the universe is
\begin{equation}
T_U\,=\,C\frac{M_U}{M_{earth}}\frac{R_{earth}}{R_{U}} T_{earth}
= 4.1\times 10^{30} K, \label{TofU}
\end{equation}
and corresponding to this temperature energy $E_T$
\begin{equation}
E_T = (3/2)kT = 8.5\times10^7 J, \label{ET}
\end{equation}
where k is Boltzmann constant, k = 1.38 $\times 10^{-27}JK^{-1}$.

From (\ref{U(b,U)}), we have the binding energy of a baryon at the
center of the universe by the whole contracted universe U(b,U) =
- $\frac{2GM_{U}m_{b}}{R}$. Using $M_U$ = $1.0\times10^{53}kg$,
$m_b = 1.67\times10^{-27}$ and R = $2.24\times10^4m$, we get
\begin{equation}
U(b,U) = -\frac{2GM_{U}m_{b}}{R} = - 9.95\times10^{11}J.
\label{U(B,U)}
\end{equation}
From (\ref{ET}), the total binding energy $E_{Tol}$ of each baryon is
\begin{equation}
E_{Tol} = E_T + U(b,U) = 8.5\times10^7 J - 9.95\times10^{11}J
= - 9.95\times10^{11}J. \label{ETol}
\end{equation}
Formula (\ref{ETol}) shows that although the universe contracts
to the small ball with radius R = 2.24$\times10^4$m, its temperature
is as high as $T_U$ = 4.1$\times 10^{30}$K, with associated its energy
$E_T$ = 8.5$\times 10^7 J$, and the baryons in the universe cannot
overcome the gravitational binding energy (-9.95$\times10^{11}$J) to
get free from the ball. It still needs more than $9.95\times 10^{11}$
J of energy for any baryon (p or n) to create the Big Bang. When the
$R_U$ contracts, the temperature is higher from (\ref{TofU}), and the
energy $E_{T}$ is higher from (\ref{ET}); but from (\ref{U(b,U)}) the
binding energy will increase with the contraction. The increased heat
(T) energy $E_T$ of the contracted universe is completely canceled by
the increased binding energy U(b,U). Thus the total binding energy
E(WU) of the whole
contracted universe with R = 2.24$\times10^4m$
and mass M = $1.0\times10^{53}$ is\\
\begin{equation}
E(WU) = E_{Tol} \times N_{b,U} = - 9.95\times10^{11}J \times \frac
{1.0\times10^{53}}{1.67\times10^{-27}} = - 5.96\times 10^{91}J.
\label{E(WU)}
\end{equation}
Therefore the total energy E(WU) produced by the contracted universe
is a negative energy, - 5.94$\times 10^{91}$J, it cannot make the
Big Bang as shown in Table 1. \\
\begin{tabular}{l}
$\qquad$Table 1. The Radius $R_U$ and the Total Energy
E(WU) of the Contracted Universe \\
\begin{tabular}{|l|l|l|l|l|l|l|l|l|l|l|}
\hline $R_U$(m),\,\, & 2.24$\times10^{-26}$ &
2.24$\times10^{-6}$ & 2.24 & 2.24$\times10^{4}$ &
2.24 $\times10^{8}$ \\

\hline $\overline{\rho}(kg/m^3)$\,(\ref{rho})\, & 2.13$\times10^
{129}$ & 2.13$\times10^{69}$ & 2.13$\times10^{51}$ & 2.13
$\times10^{39}$ & 2.13$\times10^{27}$ \\

\hline $T_U$(K),\,\,(\ref{TofU})& 4.1$\times 10^{60}$ & 4.1
$\times 10^{40}$ & 4.1$\times 10^{34}$ & 4.1$\times 10^{30}$ &
4.1$\times10^{26}$\\

\hline $E_T$(J),\,\, (\ref{ET})& 8.5$\times 10^{37}$ &
8.5$\times 10^{17}$ & 8.5$\times10^{11}$ & 8.5$\times10^7 $ &
8.5$\times10^3 $ \\

\hline  $U_{b,U}$(J),\,\,(\ref{U(B,U)}) & - 9.95$\times10^{41}$ &
- 9.95$\times10^{21}$ & - 9.95$\times10^{15}$ & - 9.95$\times10^
{11}$ & - 9.95$\times10^{7}$  \\

\hline  $E_{Tol}$(J),\,\,(\ref{ETol}) & - 9.95$\times10^{41}$ &
- 9.95$\times10^{21}$ & - 9.95$\times10^{15}$& - 9.95$\times10^
{11}$ & - 9.95$\times10^{7}$\\

\hline  E(WU)(J) (\ref{E(WU)})  & - 5.96$\times 10^{121}$ &
- 5.96$\times 10^{101}$ & - 5.96$\times 10^{95}$ & - 5.96$\times
10^{91}$ & - 5.96$\times 10^{87}$ \\

\hline
\end{tabular}%
\end{tabular}
$\qquad$ $\qquad$ $\qquad$\\
In order to show the contraction of the universe cannot produce the
Big Bang, we list the total energy E(WU) values that will be
produced by the corresponding contracted radius $R_U$ values as shown
in Table 1. In Table 1, the numbers in the parentheses of the first
column (\ref{rho}), (\ref{TofU}), (\ref{ET}), (\ref{U(B,U)}),
(\ref{ETol}) and (\ref{E(WU)}) are the numbers of formulae. Table 1
shows that although the contraction of the universe really can produce
extremely high density $\overline{\rho}(kg/m^3)$ from (\ref{rho})
and high temperature $T_U$(K) from (\ref{TofU}), at the same time
it can produce much larger binding energy $U_{b,U}$(J) from
(\ref{U(B,U)}) also. The final total result is the $E_{Tol}$(J)
from (\ref{ETol}) for each baryon.

For the whole universe, the total energy of the universe E(WU) from
(\ref{E(WU)}) is shown on the last row of Table 1. The total energy
E(WU) produced by the contracted universe can not produce the Big
Bang since the total energy is always negative. As the contracted
size of the universe becomes smaller, the total energy becomes larger
(absolutely value), and the production of the Big Bang becomes much
more difficult. The contracted universe, however, can pull the negative
energy quarks out from the quark Dirac sea to prepare the ``fuel" for
the Big Bang.

\subsection{The binding energy of a quark inside the quark Dirac
sea}
The interactions among the quarks inside the quark Dirac sea are
electromagnetic interactions, strong interactions and gravitational
interactions. The strong interactions are the strongest. It mainly
determines and holds the structure of the quark Dirac sea together.
Our purpose primarily is to find the strong binding energy of the
quark now. We do not know how to calculate it directly. Thus we
calculate the electromagnetic binding energy of the quark first,
then using strong interactions are 137 times larger than
electromagnetic ones to estimate indirectly strong bind energy of
the quark by the quark Dirac sea. Since electrons do not have strong
interactions, we omit them and only consider the quarks. We will
find the electromagnetic binding energy of the quark in the quark
Dirac sea first.
\subsubsection{Estimating electromagnetic binding energy of
a quark in the quark Dirac sea}
In the quark Dirac sea, the u-quarks have Q = 2/3 and the
d-quarks have Q = -1/3, they interact each other. This is a
many-body problem. It cannot be resolve precisely. We have to use
an approximate method. The quark Dirac sea is composed of a huge
number domains. There is a piece of body-central cubic quark
lattice in each domain as shown in Figure 1 (a). In the body-central
cubic quark lattice, each u-quark always have eight nearest neighbor
d-quarks and each d-quark always have eight nearest nearest u-quarks
as shown in Figure 2 (b). The quarks at centers have opposed electric
charges with the nearest neighbor quarks at the corners. The simplest
approximation is the nearest neighbor approximation. In this
approximation, for any quark, we only consider its eight nearest
neighbor quarks (omit all of other quarks) since other quarks with
positive charges nearby negative charges contribute to the binding
energy approximately cancel with each other. In the body-central
cubic quark lattice, for any d-quark with Q = - 1/3, there are
always eight nearest neighbor u-quarks with Q = +2/3 (see Fig. 2 (b))
and for any u-quark with Q = +2/3, there are always eight nearest
neighbor d-quarks with Q = -1/3. Thus in the nearest neighbor
approximation, for any quark (u or d), the binding energy u(e,q) =
u(e,d) = u(e,u) is
\begin{equation}
u(e,q)\,\,=\,\,u(e,d)\,\,=\,\,u(e,u)\,\,=
\,\,8\frac{KQ_{cen}Q_{cor}}{d}\,\,= \,- \,2.92\times10^{7}(J),
\label{u(e,q)}
\end{equation}
where K = 8.98755 $\times$ $10^9$ N $m^2$ $C^{-2}$, $Q_{cen}$ =
(-1/3)1.60$\times10^{-19}C$, $Q_{cor}$ = +(2/3)1.60$\times$
$10^{-19}C$ and d = ($\sqrt{3}$/2)a. Where ``a" is the lattice
constant (``a"= the Planck length = 1.62$\times10^{-35}$ m from
\textbf{Postulate III}).
Formula (\ref{u(e,q)}) shows that any quark has approximately
the same binding energy u(e,q) = u(e,u) = u(e,d) =
- 2.92$\times10^{7}(J)$.
\subsubsection{Estimating the strong binding energy of
a quark inside the quark Dirac sea}
We can not directly calculate the strong binding energy of a quark
in the QDS since the quark Dirac sea is a many body system; we have
to use an approximation method to estimate the quark strong binding
energy. Because three nearest neighboring quarks have white color
without strong interactions with other quarks. Thus we use the nearest
neighbor approximation to consider only the eight nearest quarks and
omit all other quarks. Although using the nearest neighbor
approximation, we still can not directly calculate the strong binding
energy of the quark with eight nearest neighbor quarks since this is
a many body problem. Additional the high temperature, the high density,
the high pressure and the small distance ($10^{-35}$ m) altogether
further increase the difficult of the calculation. Fortunately, we have
already found the electromagnetic binding energy (\ref{u(e,q)}) between
the quark and its eight nearest quarks. For simplicity, using strong
interactions are 137 times larger than the electromagnetic interactions,
we rough estimate the strong binding energy U(s) between the quark and
its eight nearest neighbor quarks:
\begin{equation}
U(s)\,\,=\,137u(e,q)\,\,=\,-\,4.00\times10^{9}\,(J). \label{U(s)}
\end{equation}
\subsubsection{Estimating the gravitational potential energy of
a quark in the quark Dirac sea}
Any quark in the quark Dirac sea has ``relatively infinite" quarks
around it. The ``relatively infinite" quarks have ``relatively
infinite" mass. Thus the quark has a ``relatively infinite"
gravitational potential energy U(G); it is the same everywhere
since the quark Dirac sea is homogenous and isotropic. We take
the potential energy U(G) as the energy zero point from
\textbf{Postulate VI}.
\begin{equation}
U(G)\,\,=\,\,0. \label{U(G)}
\end{equation}
Thus the gravity energy is the same anywhere, and the gravitational
field $F_G$ equals zero anywhere.
\subsubsection{Estimating the total binding energy of a quark in
the quark Dirac sea}
A quark inside the quark Dirac sea has the electromagnetic binding energy
u(e,q) = -\,2.94$\times10^{7}$(J) from (\ref{u(e,q)}) and the strong
binding energy U(s)= - 4.00$\times10^{9}$(J) from (\ref{U(s)}) as well
as the gravitational potential energy U(G) = 0 from (\ref{U(G)}).
Thus the total binding energy U(T) is
\begin{equation}
U(T)\,\, = u(e,q)\,+\,U(s)\,\,+ \,\,U(G)\,\,= - \,4.03\times 10^{9}(J)
\label{U(Tol)}
\end{equation}\\
We have already found the binding energy U(T) of a quark inside the
quark Dirac sea (\ref{U(Tol)}), and we have already calculated the
gravitational pulling energy of the quark inside the QDS by the
universe (\ref{U(b,U)}). Comparing the two kinds of energies, we can
find the condition for a quark to be pulled out from the QDS. The
condition is that the absolute value $|U(cent)|$ of the gravitational
pulling energy on the quark is larger than the absolute value $|U(T)|$
of the binding energy of the quark inside the quark Dirac sea.
\subsection{The contracted universe pulls a huge amount of quarks
and antiquarks out from the quark Dirac sea with its gravity}
Formula (\ref{U(cent)}) shows that the universe's pulling energy
of a quark in the QDS proportions to $\frac{1}{R}$. This means that
the unverse's pulling energy (absolute value) increases as the
universe's radius R decreases. Since the binding energy of a quark
in the quark Dirac sea is negative and the pulling energy of the
quark from the universe is negative too, once the pulling energy's
absolute value ($|U(center)|$) is larger than the binding energy's
absolute value ($|U(T)|$ = $4.03\times10^9$J) of the quark in the
quark Dirac sea, the quark will be pulled out from the quark Dirac
sea. Here we omit the energy $E_T$ of the temperature from contraction
of the universe (about 0.01 $\%$ from Table 1).
\begin{equation}
|U(cent)|\,\geq\,|U(T)|\,\to\,\,\frac{2GM_{0}m_{q}}{R}\,\,\geq
4.03\times 10^{9} J.
\label{U(c) = U(T)}
\end{equation}
About the initial mass $M_0$ of the universe we do not know; we
are not sure whether the dark matter and dark energy exist before
the Big Bang. For simplicity, we assume that before the Big
Bang the universe has only the observable mass of the current
universe $M_0$ = $M_U$ = $1.0\times10^{53}$kg from (\ref{Universe});
where $m_q$ = 3.80 Mev is the average mass of u-quarks ($m_u$ = 2.55)
and d-quarks ($m_d$ = 5.04) \cite{Quark mass}. Thus we have
\begin{equation}
R\,\,\leq\,\,\frac{2GM_{U}m_{q}}{4.03\times
10^{9}}\,= 2.24\times10^{4} m. \label{R}
\end{equation}

When R $\leq$ 2.24 $\times10^4$ m, the quark at the center of the
universe ball (the quark with negative energy inside the quark
Dirac sea in the vacuum) is pulled out from the quark Dirac sea
by the gravity of the universe. After the quark at the center of
the universe is pulled out from the quark Dirac sea, then: 1). The
quarks binding energy on the eight former nearest neighbor quarks
in the quark Dirac sea decreases by $\frac{1}{8}$. This is a big
amount. 2). The universe's pulling energy absolute value($\frac{
2GM_U m_q}{R}(1-\frac{r_0}{2R})$) on the eight former nearest
neighbor quarks in QDS decrease $\frac{0.866a}{2R}$ = 1.26$\times
10^{-30}$(J) than U(center) ($\geq 4.03\times10^{11}$J) since $r_0$
= $\frac{\sqrt{3}}{2}a$ (this is very small and can be omitted).
Thus the former nearest eight neighbor quarks are pulled out from
the quark Dirac sea also since the gravity force is the same as
before but the binding force decreases by $\frac{1}{8}$. Similarly,
the former nearest neighbor quarks of the former eight nearest
neighbor quarks are pulled (excited) out from the quark Dirac sea
too. The process continues $\ldots$. 3). Until the universe expands
to its size larger enough and the pulling energy (absolute value)
small enough (small than 4.03$\times10^{9}$ J) which can not pull
quark out the quark Dirac sea any more, the pulled quark process
stop. The precise number of the pulled quarks are unknown, but we
do know the number is huge and the number of pulled antiquarks is
huge also, furthermore the number of the pulled quarks equals the
number of the pulled antiquarks. The pulled huge number of quarks
and antiquarks will combine into baryons and release a huge amount
of energy to produce the Big Bang.

\section{The Quark Energy Produces the Big Bang}
When the universe contracts to the critical size with R = 2.24
$\times10^4$ m, the universe pulls a huge number of quarks and
antiquarks from the quark Dirac sea. The excited huge amount of quarks
and antiquarks are the fuel to produce the energy needed by the Big
Bang. Let us estimate how many excited quarks and antiquarks can make
the current expanding universe.
\subsection{How many excited quarks and antiquarks can make
the current expansion of the universe}
In order to know how many excited quarks and antiquarks can make
the current expansion, we have to know the whole binding energy of
the contracted universe at the critical size R = 2.24$\times10^4$ m.
In this paper, we will usually use estimated values to illustrate our
physical ideas. We do not claim to be accurate for our estimates. In
fact, some astronomical values cannot be given accurate values right
now, we have to estimate them.

\subsubsection{Estimating the whole binding energy of the universe
at critical size}
From Formula (\ref{TofU}), the contracted universe at the critical
radius R = 2.24$\times10^4$ m has $T_U$\,=\,4.1$\times10^{30}K$ .
The temperature provides energy $E_T$ = $\frac{3kT}{2}$ = 8.5$\times
10^{7}$J for a baryon from Formula (\ref{ET}). From Formula
(\ref{U(b,U)}), the pulling energy of a baryon by the universe U(b,U) =
 - 9.95$\times 10^{11}$J for a baryon. The total binding energy of a
baryon inside the universe is $E_{Tol}$ = $E_T$ + U(b,U) =
8.5$\times10^{7}$J\,-\,9.95$\times10^{11}$J = - 9.95$\times10^{11}$J.
Thus the whole universe with mass 1.0$\times10^{53}$kg has total
binding energy E(WU) = - 5.96$\times10^{91}$J from (\ref{E(WU)}).

Without energy comes from outside, the universe cannot expand.
The energy of making the Big Bang is from the excited quarks and
antiquarks. In order to know how many excited quarks and antiquarks
can produce the energy needed by the expansion of the current
universe, we have to know how much energy can be released when three
quarks combine into one baryon and three antiquarks combine into one
antibaryon as well as when one baryon-antibaryon pair annihilates.

\subsubsection{How much energy can be released when three quarks
(antiquarks) combine into a baryon (antibaryon) and one quark-antiquark
pair annihilates?}

The three excited quarks combine into a baryon and release energy,
and the three excited antiquarks combine into an antibaryon and
release the same quantity of energy. How much energy can be released
when three quarks (antiquarks) combine into a baryon (antibaryon)?

This is a very difficult problem. Directly measuring the released
energy is not possible right now. In fact physicists have tried to
separate baryons to get free quarks for many decades, so that we have
some information about the energy to separate the quarks inside the
baryon. Physicists have already used the accelerators that have the
highest energy 980 Gev in Fermilab (Tevatron 1978-present and
Tevatron II 2001-present) for many years. No individual free quark
has been found \cite{Free Quark}. This kinds of accelerators (980
Gev) cannot free the quarks from baryons. This means that the energy
(980 Gev) cannot separate the quarks in baryons. The cosmic rays with
much higher energy than the energy (980 Gev) steel cannot separate
the quarks in baryons also. Thus we estimate the released energy of
three quarks combining into a baryon is much larger than the 980 Gev.
If we want to separate a baryon with 100$\%$ possibility, we need
much higher energy than the energy (980 Gev). Therefore we roughly
estimate the three quarks binding energy E(release) (=E(melt) that
can separates baryon into quarks and gluons) is about
\begin{equation}
E(release)\,\,\, > \,\,\, 980 \,Gev. \label{B-Bind}
\end{equation}
Formula (\ref{B-Bind}) show that three quarks combine into a baryon to
release an energy (much more than 980 Gev ) and three antiquarks combine
into an antibaryon to release an energy (much more than 980 Gev) also.
Then the baryon-antibaryon pair annihilates back the quark Dirac sea
and release 1878 Mev (two baryons with energy 939+939 Mev) energy at
least. The total released energy $E(b+\bar{b}+b\bar{b})$ of the
excited three quarks and the three antiquarks as well as a baryon-
antibaryon pair will be
\begin{equation}
E(b+\bar{b}+b\bar{b}) \gg 1962\, Gev = 3.14\times10^{-7}J. \label{1962}
\end{equation}
Formula (\ref{1962}) shows that three quarks combine into a baryon
and the three antiquarks combine into an antibaryon as well as a
baryon-antibaryon pair annihilates together to release much more than
3.14$\times10^{-7}$J energy.
\subsubsection{How much energy can be released from a pair q$\bar{q}$
annihilation?}
How much energy can be released from a pair q$\bar{q}$ annihilation?
From Formula (\ref{1962}), the three quarks combine into a baryon and the
three antiquarks combine into an antibaryon as well as a baryon-antibaryon
pair annihilates together to release more than 3.14$\times10^{-7}$J energy.

Since a baryon is composed of three quarks and an antibaryon is composed of
three antiquarks, one baryon-antibaryon  pair is equivalents to three
quark-antiquark pairs. Thus we guess that one quark-antiquark pair
annihilation can releases energy, from Formula (\ref{1962}), as
\begin{equation}
E(q\bar{q})\,=\,\frac{E(b+\bar{b}+b\bar{b})}{3} > 654 Gev =\,
1.05\times10^{-7}J. \label{E(q-q)}
\end{equation}
Formula (\ref{E(q-q)}) shows that one quark-antiquark pair
annihilation can release energy $E(q\bar{q})$ much larger than
$1.05\times10^{-7}$J.

\subsubsection{How many excited quarks are need to make the current
expanded universe?}
From Formula (\ref{E(WU)}), we know that the whole observable
contracted universe with mass M = $1.0\times10^{53}$ kg, at R =
$2.24\times 10^4$ m has binding energy E(WU) = - 5.96$\times
10^{91}$J. Formula (\ref{E(q-q)}) gives that one quark-antiquark
pair annihilation can release energy $E(q\bar{q})$ much larger than
$1.05\times10^{-7}$J. Thus we need
\begin{equation}
N(q) = N(\bar{q}) < \frac{5.96\times 10^{91}J}{1.05\times10^{-7}J}
= 5.68\times10^{98}. \label{N(q)}
\end{equation}
The excited quark number N(q) equals the excited antiquark
number $N(\bar{q})$ is much smaller than 5.68$\times10^{98}$. The
number is the needed number of quarks and antiquarks which can release
enough energy to push the observable matter of the universe
($1.0\times10^{53}$ kg) to infinity with no kinetic energy.

This matter ($1.0\times 10^{53}$ kg) is the observable matter of the
universe. It does not include the dark matter and dark energy. The
dark matter and the dark energy have 96$\%$ percent of the total
matter of current universe, while the observable matter is 4 $\%$ of
the total matter of current universe. If we push all matter
(including dark matter and dark energy) to infinity, we need
more excited quarks N(q,D) and antiquarks N($\bar{q},D)$):
\begin{equation}
N(q,D) = N(\bar{q},D) < \frac{5.96\times 10^{91}J}{1.05\times10^{-7}J}
\times \frac{100}{4} = 1.42\times10^{100}. \label{N(q,D)}
\end{equation}
We do not know how many excited quarks are needed in the Big Bang
from the starting expansion to the present state, but we can guess that
the real needed number N(Need) is smaller than the number N(q,D) $\leq$ $1.42\times10^{100}$ since the universe has not expanded to infinite yet,
and it will be larger than N(q) = 5.68$\times10^{98}$ since the total
mass of the current universe is much larger (24 times) than the observable
mass of the universe, we guess it may more than N(q)

\begin{equation}
5.68\times10^{98}\,<\,N(Need)\,<\,1.42\times10^{100}.
\label{N(Need)}
\end{equation}
Formula (\ref{N(Need)}) gives the needed excited quark number range
5.68$\times10^{98}$--1.42$\times10^{100}$. A huge number of the excited
quarks and antiquarks are pulled out from the quark Dirac sea in a very
short time.

\subsection{Loading up the fuel (quarks and antiquarks)
for the Big Bang}
When the universe contracts to its critical size R = 2.24$\times10^4$ m,
three processes start:

1). The universe pulls quarks and antiquarks out from the quark Dirac
sea; adding the mass to the universe (and the pulling force of the
universe) and decreasing the binding energy of the nearest former
remaining quarks in the quark Dirac sea; so that the pulling quark
process continues progressing.

2). The excited (pulled) quarks (antiquarks) combine into baryons
(antibaryons) and release energy since three reasons: First, under
usual temperature, ordinal pressure and normal gravitational field,
the ``melting-point" of baryon (the temperature at which baryons can be
separated into quarks and gluons) is unknown high (much high than 980
Gev, from (\ref{B-Bind})). Second, although the contracted universe with
R = 2.24 $\times10^4$ m has very high temperature (4.1$\times10^{30}$K
$\to$ $E_T$ = 8.5$\times10^7$ J) which may melt baryons and obstruct
quarks combine into baryons. There is, however, the gravitational energy
U(b,U) = - 9.95$\times10^{11}$J which has canceled the $E_T$ and help
the quarks to combine baryons. Third, the very high density ($\rho$ =
2.13$ \times10^{39}$kg) and the very strong gravitational field of the
contracted universe produce very high pressure to help that the quarks
combine into the baryons. In fact, there is really a huge amount energy
to produce the Big Bang, this energy cannot come from the atoms and nuclei,
it is only comes the quarks. The quark energy is really produced and
really makes the Big Bang.

3). The released energies start and accelerate the expansion
of the universe. The universe's size (R) becomes larger and larger, and
the universe pulling force becomes smaller and smaller. Until the size
reaches a new critical value $R^*$ and $M^*_0$, the universe pulling
energy (absolute value) from Formula (\ref{U(b,U)}) will be smaller
than the binding energy (absolute value) of the quark Dirac sea in
Formula (\ref{U(Tol)}), and the universe cannot pull out quark any more.
\begin{equation}
|\frac{2GM^*_0m_q}{R^*} (1-\frac{r_0}{2R^*})| < |U(T)|\,=
4.03\times 10^{9}(J)
\end{equation}
where $m^*_0$ is the mass of the universe at R = $R^*$; it includes
the observable mass and the mass of the excited quarks (and the excited
antiquarks).

In this period, the main job is loading the fuel for the
Big Bang. The fuel is the excited quarks and antiquarks that are
pulled out from the quark Dirac sea by the contracted universe using
its gravity. From Formula (\ref{N(Need)}), the total excited quark
number is about 5.68$\times10^{99}$--1.42$\times10^{100}$. Since the
process is incredibly fast, the time of the process is incredibly
short. When the process of the universe pulling quarks is finished,
this means the loading fuel period is over, and a new period is
coming.

\subsection{The explosive accelerated expansion of the universe}
After the loading of the fuel, the universe has a huge amount of fuel
(excited quarks and antiquarks). The excited quarks and antiquarks are
expanding with the original matter of the universe, so that the quarks
and antiquarks broadcast everywhere in the universe. In order to
understand the real state of the universe, we can estimate the
broadcasted quark density. We assume that when the universe expands
to radius $R^*$ = 100$R = 2.24\times10^6$m and the excited quark
number equals half N(Need) in Formula (\ref{N(Need)}), the volume of
the universe V = $\frac{4\pi}{3}$(2.24$\times 10^6m)^{3}$ = $4.71
\times10^{19}m^{3}$. Using N(q,D) =1.42$\times 10^{100}$ from
(\ref{N(q,D)}), we get
\begin{equation}
\rho(upper)= 0.5\frac{N(q,D)}{4.71\times10^{19}m^{3}}\,
=\,1.51\times10^{80}m^{-3}.
\label{rho(upper)}
\end{equation}
Using N(q) = 5.68$\times10^{98}$ from (\ref{N(q)}), we get
\begin{equation}
\rho(lower)= 0.5\frac{N(q)}{4.71\times10^{19}m^{3}}\,
=\,6.03\times10^{78}m^{-3}.
\label{rho(lower)}
\end{equation}
Formula $(\ref{rho(upper)})$ and Formula $(\ref{rho(lower)})$ show the
density up limit and down limit. The density is very high, under the
strong interactions, the excited quarks and antiquarks have explosive
quark reactions. In order to precisely express the meaning of ``nearby",
we define: if two particles are inside the strong interactions range
(about $10^{-15}$m), then we call them nearby particles. If a quark has
a nearby antiquark, they will annihilate and release energy; if three
quarks (uud or udd) are nearby with each other, they will combine into
a baryon and release energy; if three antiquarks $(\overline{uud}$, or
$\overline{udd})$ are nearby with each other, they will combine into
an antibaryon and release energy; if an antibaryon has a nearby baryon
they will annihilate and release energy. Since there are a huge number
of quarks and a huge number of antiquarks, the above processes will
continuously explosive progress, and the speed of the expansion of the
universe will accelerate faster and faster until the nearby
quark-antiquark pairs run out, the nearby quarks run out, the nearby
antiquarks run out, then the explosive accelerating engine stop.
Although the explosive acceleration time is very short, a huge
quantity of energy E(Out) has already released in this period. The
universe in this process is like the explosion of a bomb; but the
energy of the universal Big Bang is much larger than a hydrogen
bomb.

In order to see this case, we can estimate the released energy E(Out)
of the quarks and the antiquarks. We assume 95 percent of the total
needed quarks and antiquarks have been used in the period. From
Formula (\ref{N(Need)}) and Formula (\ref{1962}), the released energy
range Range(E(Out)) is
\begin{equation}
Range(E(Out)) = 0.95\frac{N(need)}{3}\times E(b+\bar{b}+b\bar{b}) =
5.65\times10^{91}J \to 1.41\times10^{93}J. \label{E(Out)}
\end{equation}
Formula (\ref{E(Out)}) gives the released energy region 5.65$\times10^{91}$J---1.41$\times10^{93}$J. This is a huge energy
that is enough to overcome the gravity of the universe and to
produce the Big Bang. Explaining the origin of the energy of the
Big Bang is the main purpose of this paper.

After the fast expansion dilutes the distribution of the quarks and
antiquarks, the density of the excited quarks and antiquarks becomes
much smaller. A very slowly accelerated period is coming.

\subsection{A very slow acceleration and a very fast expansion period}

At the end of the explosive accelerated expansion period, the quark
(or antiquark) density has already become very small. Although the
density is very small, the total excited quark number is still
huge (about 5 percent of total needed quarks 5.68$\times 10^{98}$--
1.42$\times 10^{100}$ = 2.83$\times 10^{97}$--7.10$\times 10^{99}$).
Although the acceleration is very small, the speed of the expansion
is large. The large expansion of the universe dilutes the distribution
of the excited quarks, as the time passes, the density of the quarks
and antiquarks becomes smaller and smaller; the released quark energy
becomes smaller and smaller; as a result, the acceleration becomes
smaller and smaller. Maybe the acceleration is difficult to be found.
We do not know how long this process will take.

In fact, many physicists have searched for free quarks for more than
thirty years, no free quark has been found \cite{Free Quark}. This
means there is no free quark. Thus the observable acceleration of the
expansion of the universe now is not from the quarks. The quark
acceleration has already finished. If the observable acceleration
is really come from the quarks, today observed acceleration may be
produced long time ago since the light needs time to travel to the
earth. Because of the current experiments have not found any
individually quark.

\subsection{The future of the universe}
The current universe is in a state of slowly accelerating expansion.
The future of the universe is uncertain. It depends on how long the
acceleration can keep up. If the acceleration can keep up forever,
then the universe will expand forever. If the acceleration can keep
up for only a finite time, the gravity of the universe will stop the
expansion and the gravity will recontract the universe again. Thus,
the universe will be cyclic: the universe will undergo a series of
cycles, each beginning with a big bang and ending with a big crunch
\cite{Baum} \cite{Bounce}.

\section{Discussion}

Now that the quark Dirac sea really exists in the vacuum,
why has it not been discovered?

The quark Dirac sea has not been discovered because of at least
six reasons:

First, because the lattice constant of the quark Dirac sea
a = ($1.62\times10^{-35}m$) and the sizes of the domains are much
smaller than the limit of the distance scale of the standard model
$10^{-18}m$ \cite{Standard}. The domains are distributed completely
randomly in positions and in directions of symmetry axes of the
lattices. The QDS appears as a homogeneous and isotropic as well as
``continuous" space.

Second, there are no net electric charge, no net color charge,
no the gravitational potential and no the gravitational field ($V_G$
= 0 and $F_G$ = 0 from \textbf{\textbf{Postulate VI}}) at any physical
point inside the quark Dirac sea. Thus, physicists cannot use electric
or strong or gravitational interactions to discover the neutral quark
Dirac sea. The QDS appears as a completely ``empty" space.

Third, the quarks (u-quarks or d-quarks) in the quark Dirac sea
are bound by the strong attractive forces and Coulomb's attractive
forces with energy (4.03$\times$ $10^{9}(J)$ $(\ref{U(Tol)})$). It is
a super-strong structure. Even nuclear fusion energy (about 10 Mev =
$10^{-12}$ J) is not enough to change the quark Dirac sea for
detection. Thus it looks like it can never be changed. No change
means no chance to be discovered.

Fourth, from Formula (\ref{U(b,U)}) we can estimate the pulling energy
of a baryon at the center of the earth, the Sun and the Milky Way
with R = $10^3$m. We get $E_{earth}$ = - $\frac{2GM_{earth}m_b}{R}$
= - 1.33 $\times10^{-15}$J; $E_{Sun}$ = - $\frac{2GM_{Sun}m_b}{R}$
= 4.43 $\times10^{-10}$J; $E_{Milky Way}$ = - $\frac{2GM_{Milky
Way}m_b}{R}$ = - $\frac{2G\times5.8\times10^{11}M_{Sun}m_b}{R}$ =
- 5.43$\times10^2$J. These energies represent the pulling energies of
planets, stars and galaxies. All the pulling energies are much smaller
than the banding energy (- 4.03$\times$ $10^{9}(J)$ $(\ref{U(Tol)})$)
of the quark Dirac sea. All celestial bodies pulling energy can be
omitted. The quark Dirac sea looks as ``empty" since a huge galaxy
contracted to very small size with R = 1000m cannot pull a quark out
from it.

Fifth, the ideal lattices do not scatter a particle moving inside
it \cite{QM}. The lattices inside the domains of the QDS are all the
ideal lattices. The QDS appears as a no scattering space.

Sixth, the quark Dirac sea is a ``relatively infinite" homogeneous
and isotropic system. All scientific observations, experiments and
research occurs within the quark Dirac sea. This (``relatively
infinite" system) has limited the scientists from comparing the
current case with the other case without the quark Dirac sea to
find the quark Dirac sea.

Despite, the above obstacle, in fact scientists have already
discovered some clues of the quark Dirac sea. For example, high energy
physical experiments have discovered large numbers of particle pairs
(e$\overline{e}$, p$\overline{p}$ and n$\overline{n}$) created out from
the vacuum and annihilated back to the vacuum. This is a clue that the
quark Dirac sea exists. If there is no quark Dirac sea, where do these
particles come from?

\section{Conclusions}
1. The contraction of the universe cannot produce the Big Bang as
Table 1 has shown. We have to expand the original Dirac sea (including
electrons only) to the quark Dirac sea (QDS) including negative energy
u-quarks and d-quarks as well as electrons in the vacuum.

2. There is a huge number of domains of the body-central cubic
quark lattice with the size much smaller than $10^{-18}$m. They are
completely randomly distributed in positions and directions of the
symmetry axes of the lattices in the domains fully over the QDS. In
the lattice, each primitive cell contains two quarks (u and d), the
nearest three cells contain an electron. Its lattice constant is the
Planck length ``a" = 1.62$\times 10^{-35}m$.

3. The QDS is homogeneous, isotropic and ``continuous". There is no
net electric charge, no net color charge, no gravitational potential
($V_{QD}$ = 0) and no gravitational field ($F_G$ = 0) at any physical
point in the QDS. The particles are moving in the QDS likes they are
moving in a stable and non-scattering space. It is a warehouse of
$e\bar{e}$, $u\bar{u}$, $d\bar{d}$, $p\bar{p}$ and $n\bar{n}$,$\ldots$,
pairs. Therefore the quark Dirac sea is a perfect model of the vacuum.

4. The contracted universe pulls out a huge number of quarks and
antiquarks from the QDS by its gravity at its critical size. This is
a necessary and sufficient condition for the Big Bang.

5. The huge number of excited quark-antiquark pairs annihilate(s)
back to the QDS and release a huge amount of energy; the huge number
of excited quarks combine into baryons and release a huge amount
of energy; the huge number of antibaryons combine into antibaryons
and release a huge amount of energy; then the B$\bar{B}$ pairs
annihilate back to the QDS and release a huge amount of energy also.
Together, these huge quark energies make the Big Bang.

6. The Big Bang theory depends on two major assumptions: the
universality of physical laws and the Cosmological Principle. The
Cosmological Principle states that on large scales the universe is
homogeneous and isotropic. Thus, the homogeneous and isotropic quark
Dirac sea provides a firmly physical foundation for the Cosmological
Principle. So that, the quark Dirac sea provides a physical firm
foundation for the Big Bang theory.

7. The present universe is expanding from a finite size universe
with a radius R = $2.24\times10^4$ m, not from the singularity
with an infinite density and temperature.

8. It provides a strong support of the quark model that only the quark energy can
make the Big Bang.

9.

8. In astronomy there are many important physical quantities that
cannot be precisely measured now. We have estimated many quantities
to illustrate our physical ideas. We are not sure the accuracy of
our estimates. They are not the results of precise measurements or
accurate theory deduction. They are only phenomenological estimates.

\begin{center}
\bigskip \textbf{Acknowledgments}
\end{center}
 I sincerely thank Professor Robert L. Anderson for his
valuable advice. I acknowledge my indebtedness to Professor
D. P. Landau for his advice and help also. I would like to express my
heartfelt gratitude to Professor H. Y. Guo and Professor Zhan Xu.
I thank Dr. Xin Yu very much for his longtime help. I thank Professor
L. A. Magnani for his advice. I also thank my son Yong-Kai for drawing
the figures.

,

\end{document}